\documentclass[preprint2,longabstract]{aastex}

\usepackage{enumerate}
\usepackage{color}
\def\msun{M$_{\odot}$}
\def\rsun{R$_{\odot}$}

\def\mdot{$\dot M$}

\def\it{\sl}
\def\degs{\ifmmode ^{\circ}\else$^{\circ}$\fi}
\def\amin{\ifmmode ^{\prime}\else$^{\prime}$\fi}
\def\asec{\ifmmode ^{\prime\prime}\else$^{\prime\prime}$\fi}
\def\degs{\ifmmode ^{\circ}\else$^{\circ}$\fi}
\def\amin{\ifmmode ^{\prime}\else$^{\prime}$\fi}

\def\eqalign#1{\null\,\vcenter{\openup1\jot \m@th
   \ialign{\strut\hfil$\displaystyle{##}$&$\displaystyle{{}##}$\hfil
   \crcr#1\crcr}}\,}
\usepackage{color}

\newcommand{\sgr}{V1082\,Sgr}

\def\apgt{\ {\raise-.5ex\hbox{$\buildrel>\over\sim$}}\ }
\def\aplt{\ {\raise-.5ex\hbox{$\buildrel<\over\sim$}}\ }

\slugcomment{Not to appear in Nonlearned J., 45.}

\shorttitle{Bottleneck accretion in long-period, magnetic  pre-CVs.}
\shortauthors{G. Tovmassian et al.}

\begin{document}


\title{LONG-ORBITAL-PERIOD PREPOLARS CONTAINING EARLY K-TYPE DONOR STARS. BOTTLENECK
ACCRETION MECHANISM IN ACTION}


\author{G. Tovmassian\altaffilmark{1}, D. Gonz\'alez--Buitrago\altaffilmark{1}, S. Zharikov\altaffilmark{1}}
\affil{Instituto de Astronom\'{\i}a, Universidad Nacional Aut\'onoma
de M\'exico, Apartado Postal 877, Ensenada, Baja California, 22800 M\'exico}
\email{gag, dgonzalez, zhar@astro.unam.mx}
%\author{K.~Mukai\altaffilmark{2,3}}
%\affil{CRESST and X-Ray Astrophysics Laboratory, NASA Goddard Space Flight  Center, Greenbelt, MD 20771, USA}
%\affil{Department of Physics, University of Maryland, Baltimore County, 1000 Hilltop Circle,Baltimore, MD 21250, USA}
%\email{aastex-help@aas.org}
%\author{F.~Bernardini\altaffilmark{4,5}}
%\affil{Wayne State University, 666 W. Hancock Street, Detroit, MI, USA}
%\affil{New York University Abu Dhabi, Abu Dhabi, United Arab Emirates}
%\and
%\author{D.~de Martino\altaffilmark{5}}
%\affil{INAF $-$  Osservatorio Astronomico di Capodimonte, Salita Moiariello 16, I-80131 Napoli, Italy}
%\and
\author{D. E. Reichart\altaffilmark{6}, J. B. Haislip\altaffilmark{6}, K. M. Ivarsen\altaffilmark{6},  A. P. LaCluyze\altaffilmark{6}, J. P. Moore\altaffilmark{6}}
\affil{Department of Physics and Astronomy, University of North Carolina at Chapel Hill, Campus Box 3255, Chapel Hill, NC 27599, USA}
\and
\author{A. S. Miroshnichenko\altaffilmark{7}}
\affil{Department of Physics and Astronomy, University of North Carolina at Greensboro, Greensboro, NC 27402-6170, USA}

\begin{abstract}
We studied two objects identified as a cataclysmic variables (CVs) with periods exceeding the natural boundary for Roche-lobe-filling zero-age main sequence (ZAMS) secondary stars. We present observational results for V1082\,Sgr with a 20.82 hr orbital period, an object that shows a low luminosity state when its flux is totally dominated by a chromospherically active K  star with no signs of ongoing accretion. Frequent accretion shutoffs, together with characteristics of emission lines in a high state, indicate that this binary system is probably detached,  and the accretion of matter on the magnetic white dwarf takes place through stellar wind from the active donor star via coupled magnetic fields. Its observational characteristics are surprisingly similar to V479\,And, a 14.5 hr binary system. They both have early K-type stars as a donor stars.  We argue that, similar to the shorter-period prepolars containing M dwarfs,  these are detached binaries with strong magnetic components. Their magnetic fields are coupled, allowing enhanced stellar wind from the K star to be captured and channeled through the bottleneck connecting the two stars onto the white dwarf's magnetic pole, mimicking a magnetic CV. Hence, they become interactive binaries before they reach contact. This will help to explain an unexpected lack of systems possessing white dwarfs with strong magnetic fields among detached white$+$red dwarf systems. 

\end{abstract}



\section{Introduction}

Cataclysmic variables (CVs) are semidetached binary systems consisting of a red star filling its corresponding Roche lobe and losing matter to a white dwarf companion \citep{1995CAS....28.....W}.  CVs with periods over 10 hr are rare, which is a natural consequence of the requirement for a (nearly) main sequence donor star to match its  Roche lobe size \citep{2012MmSAI..83..505R}. CVs with longer periods should contain late-type stars that have departed  the zero-age main sequence (ZAMS) in order to comply with the latter condition.   

Relevant to this study is a small group of objects that were thought to be low-accretion-rate polars, that is, magnetic CVs with an accretion rate of $\sim10^{-13}$\,\msun\,yr$^{-1}$\ \citep{2002ASPC..261..102S}. 
However, recently it has  been argued that these are in fact a detached pair of white and red  dwarf stars \citep{2005ASPC..330..137W,2005ApJ...630.1037S}.  A model was proposed according to which the magnetic white dwarf accretes matter captured from the wind of a magnetically active M dwarf. Hence they were dubbed  prepolars containing M-dwarf companions, which  failed to achieve Roche lobe contact in their post-common-envelope evolution. The theoretical basis for a model in which the accretion is fueled by a stellar wind from the M dwarf and  collected through interlocked magnetic fields of binary components was proposed earlier by  \citet{1994MNRAS.268...61L,1995MNRAS.276..255L}.

Here we present a new study of \sgr, an object with an extremely long period and identified as a CV.  It shows a number of outstanding observational features in a wide range of wavelengths. 
\sgr\ was discovered by \citet{1998A&AS..131..119C}.   \citet[][]{2010PASP..122.1285T}  found  absorption lines from a K-type star in its optical spectrum and using them  determined the orbital period  of 0.868 days. They tentatively  identified \sgr\ as nova-like.
\citet{2013MNRAS.435.2822B} conducted an extensive X-ray observational study, revealing  that \sgr\ is a highly variable  source,  with variations on a wide range of timescales, from hours to months. They found  that the X-ray spectrum is similar to a magnetic CV.
The object shares many  properties with V479\,And \citep{2013A&A...553A..28G}.  Taking into account that there are only a few objects identified as CVs in that orbital period range, we dwell on their similarities to understand the  underlying reasons.    
We discuss both objects  in this paper and develop a qualitative model to explain them.

\begin{table}
\caption{Log  of photometric observations.}
\begin{tabular}{l|cccl} %\hline \\ %[0.1pt]
%&  \multicolumn{4}{c}{Spectroscopy / 2.1m}   &         \\[0.1pt]
%\hline \\[0.1pt]
%Date  & Exp.  &   FWHM & Range        & Total      \\
%           &     sec         &         \AA                         &       \AA     &   hours \\[1pt]   \hline \\[0.1pt]
%19/02/12   &  600 &  2.05 & 4250-5500     &   2.0     \\
%22/07/12   &  600 &  2.05 & 4750-6000     &   2.0        \\
%23/07/12   & 600  &  2.05 & 4750-6000    &   2.0      \\
%24/07/12   & 600  &  2.05  & 4750-6000   &  2.0     \\
%25/07/12   & 600  &  2.05  & 4250-5500     & 0.3       \\
%27/07/12   & 600  &  2.05  & 4250-5500     & 1.0 \\
%28/06/13   & 600  &  2.05  & 4500-5750    & 0.5        \\
%29/06/13   & 600  &  2.05  & 4500-5760     & 0.6      \\
%07/08/13   & 600  &  4.11  & 3690-7000    & 0.8      \\
%08/08/13   & 600  &  4.11  & 3690-7000    & 0.5       \\
%09/08/13   & 600  &  4.11  & 3690-7000     & 0.5       \\
%10/08/13   & 600  &  4.11  & 3690-7000     & 0.6     \\
%11/08/13   & 600  &  4.11  & 3690-7000      &1.8     \\
%12/08/13   & 900  &  2.05  & 4500-5750     &1.5     \\
%13/08/13   & 600  &  2.05  & 4540-5750     &0.8     \\
%14/08/13   & 720  &  2.05  & 4500-5650     &2.8     \\
%15/08/13   & 600  &  2.05  & 4400-5650     &1.3     \\
%16/08/13   & 600  &  2.05  & 4400-5650    & 1.8     \\[1pt]
%\hline
%\hline \\[0.1pt]
%&\multicolumn{4}{c}{UV, optical  }   &           \\[1pt]
\hline  \hline\\
Date  & Exp.  &   Filter & Telescope        & Total      \\
           &     s         &            &     Instrument        &   (days) \\ [1pt]   \hline \\[0.1pt]
19/08/08   & 327 & UVW2 & {SWIFT}    & 0.3 \\
20/08/08   & 825 & UVM2 & {SWIFT}    & 3.7 \\
28/08/08   & 514 & UVM2 & {SWIFT}    & 0.5 \\
22/02/14   & 286 & UVM2 & {SWIFT}     & 0.7 \\
01/03/14   & 1160 & UVW2 & {SWIFT}     & 1.7 \\
08/03/14   & 1214 & UUUU  & {SWIFT}  & 1.7 \\
15/03/14   & 744 & UVW1 & {SWIFT}    & 1.8 \\
22/02/14   & 1636 & UVM2 & {SWIFT}     & 0.3 \\
01/03/14    & 1636 & UVW2 & {SWIFT}     & 0.3 \\
08/03/14    & 1636 & U  & {SWIFT}   & 0.3 \\
15/03/14   & 1636 & UVW1 & {SWIFT}    & 0.3 \\
06/13                 &  180          &    V       & APOGEE    &    19      \\
07/13                 &  180          &    V        & APOGEE   &    18      \\
08/13                 &  180          &    V        & APOGEE   &    15      \\
09/13                &  180           &    V        & APOGEE    &    14      \\
10/13                 &  180           &   V        & APOGEE    &    08      \\
11/11/13                        & 180           &   V         & APOGEE    &    01  \\ \hline
\end{tabular}
%\begin{tabular}{l}
%$^*$ Observed by {\em SDSS}. \\
%$^\star$ {Observed by {\em GALEX}. ObjID \# 6372252849676485379} \\
%$^\dagger$ {Observed by {\em SWIFT/UVOT} OBS\_ID \# 00031252 } \\
%\end{tabular}
\label{tab:log}
\end{table}

\section[]{Observations}
\subsection{Spectroscopy.}
Time-resolved spectroscopy of \sgr\  was performed with the 2.1 m telescope of the Observatorio Astron\'omico Nacional\footnote{http: www.astrossp.unam.mx}
at San Pedro M\'artir, Baja California, M\'exico (OAN SPM) in 2012 and 2013  with the Boller \& Chivens spectrograph, using a 600 and 1200 grooves mm$^{-1}$ grating with a $15\,\mu m \, 2048\times2048$\,pixel Marconi\,2 CCD, with spectral resolutions  of 4.1\,\AA \, and 1.8\,\AA,  respectively.
The standard long-slit reduction of the data with a variance weighting extraction was made using IRAF\footnote{IRAF is distributed by the National Optical Astronomy Observatory, which is operated by the Association of Universities for Research in Astronomy (AURA) under cooperative agreement with the National Science Foundation.}
procedures after applying bias subtraction. Only cleaning cosmic rays, which are abundant on 1200 s exposures,  were made with the external task {\it lacos} \citep{2001PASP..113.1420V}. 
 
The wavelength calibration was made with the help of an arc lamp taken every 10th exposure. The   Boller \& Chivens spectrograph, made originally for photographic plates, was later adopted for the CCD camera, a much heavier detector. That introduces strong flexes and significant shifts in the wavelength's zero point,  which we usually correct by using a strong sky line in each spectrum. At moderate zenith heights, those shifts can reach $1-1.5$\AA,  and the transition from exposure to exposure is smooth, so a reasonable correction is attainable. But because \sgr\ is a southern object, it remains quite low in SPM (altitude $< 35\degs$), even when the object passes through the meridian. As a result, we had severe problems with wavelength calibration, and a fraction of our spectra were worthless for radial velocity (RV) studies.

The spectra of the object  were flux calibrated using spectrophotometric standard stars observed 
during the same night. The instrument cannot automatically rotate the slit to the corresponding parallactic angle,  and for simplicity we  routinely used an E--W slit orientation. In a majority of observations the slit width was kept narrow ($180\,\mu$m = 2 arcsec) for better resolution. These two factors make a correct flux calibration difficult. But in 2014's observational run  we acquired a couple of low  and another few of higher (300 and 1200 grooves mm$^{-1}$) resolution  spectra, FWHM=8.5 and 2.2 \AA,  with a wide $350\,\mu$m slit  to circumvent this flux calibration problem. A number of K-type spectral standards from \citet{2007MNRAS.374..664C} were observed along with the object with the same instrumental settings. Two selected K2\,IV  stars, BSNS\,104 and HD\,197964, are presented here as spectral identification templates. We  used the {\it xcsao} procedure in IRAF to cross-correlate the observed spectra with a standard K star in the $\lambda 5150-5850$\,\AA\ range in order to measure the RVs of the absorption lines. A variety of standards from K2 to K4 were used, with no significant differences in the obtained velocities. 

The emission line parameters reported in this paper were measured  by fitting a single Lorentzian  because the line profiles are best described by this function. We tried  single-  and  double-Gaussian methods \citep{1980ApJ...238..946S} to determine  the RVs. The latter is  designed to estimate the orbital motion of the  accreting star  by measuring the velocity of the inner parts of the disk.   Regardless of the method, the  resulting  RV  values  are similar. 

We  make extensive use of spectra  kindly made available to us by John Thorstensen  \citep[referred to here as JT spectra;][]{2010PASP..122.1285T}.

\begin{figure}[t]
\centering
%\setlength{\unitlength}{1mm}
%\resizebox{11cm}{!}{
%\begin{picture}(100,70)(0,0)
%\put (0,0)  { \includegraphics[width=7cm,  clip]{GB_2ndry_spectrum}}
%\end{picture}}
 %\includegraphics[width=8.5cm, bb=5 20 580 560, clip]{GB_photometry}
  \includegraphics[width=8.5cm,  clip]{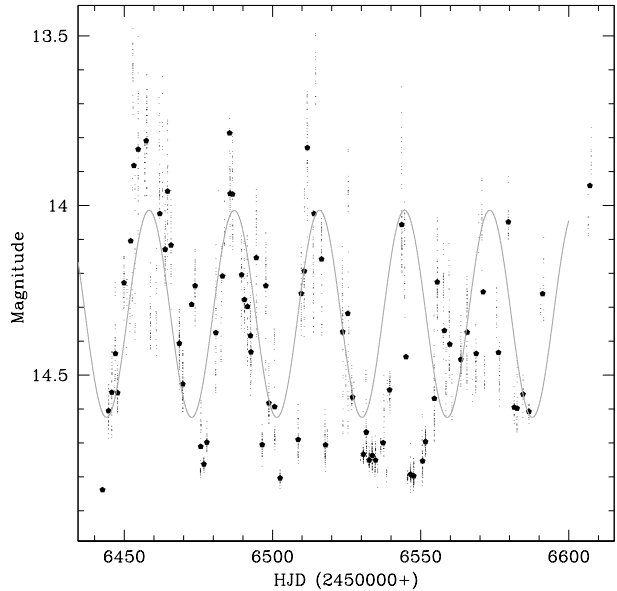}

  \caption{ The V-band light curve of \sgr\  over half a year. The individual measurements are plotted as tiny dots, and filled pentagons correspond to a nightly average magnitude.  The variability may have a periodic component at around~29 days, which is overplotted as a sine curve.   }
  \label{fig:lc}
\end{figure}

\subsection{Photometry}

Time-resolved V-band photometry was obtained using the 0.41 m  Ritchey--Chr\'etien telescope at the  Panchromatic Robotic Optical Monitoring and Polarimetry Telescopes (PROMPT) at  Cerro Tololo Inter-American Observatory (CTIO) in Chile, with the apogee camera that make use of E2V CCDs. The log of photometric observations is also given in Table~\ref{tab:log}.  The data were reduced  with IRAF,  and the images were corrected for bias and flat fields  before aperture photometry.  Flux calibration was performed using  secondary standard stars from the same field.

\begin{figure*}[th]
\setlength{\unitlength}{1mm}
\resizebox{11cm}{!}{
\begin{picture}(100,80)(0,0)
\put (0,0)  { \includegraphics[width=15cm,  bb=20 250 570 630, clip=]{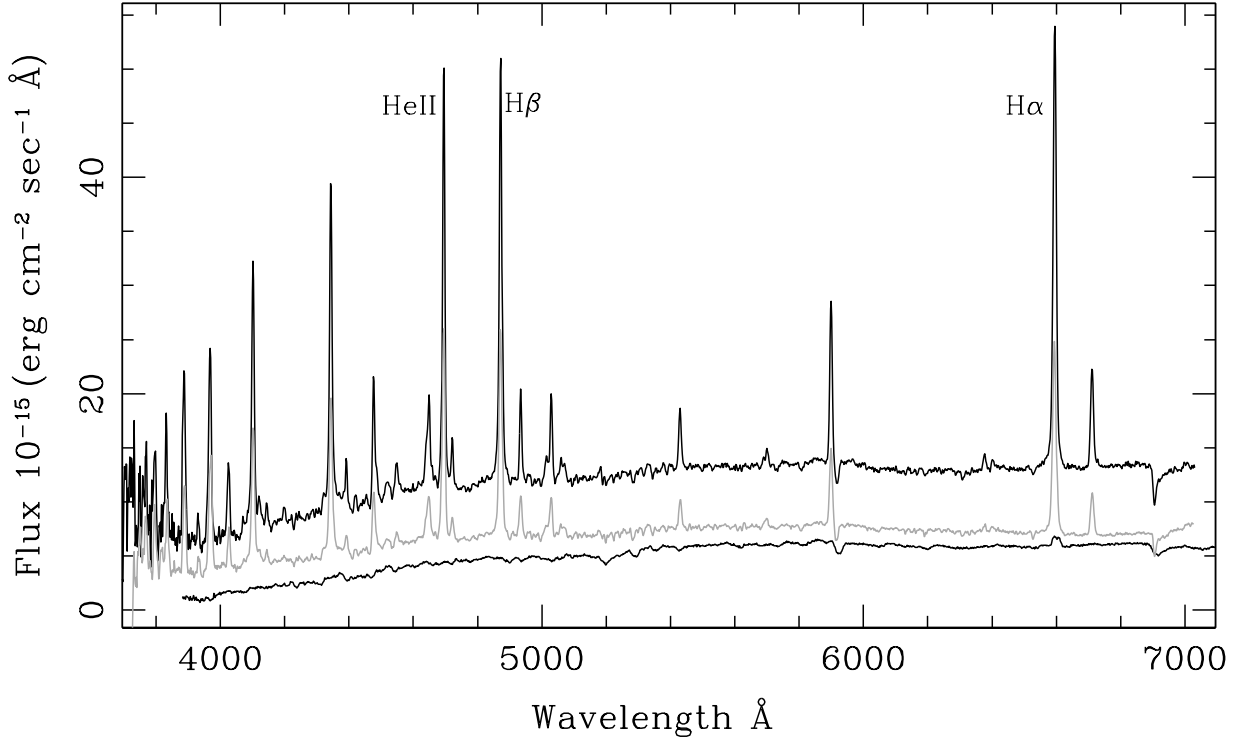}}
\end{picture}}
  \caption{ The spectra of \sgr\ in a high, intermediate, and low states. In the  high state it appears as a typical CV. In the low state, emission lines vanish altogether, but there is weak H$_\alpha$ emission. 
  }
  \label{fig:spec}
\end{figure*}

\subsection{UV and X-ray observations}

\sgr\  was observed with the {\sl Swift} telescope as a  target of opportunity (ToO target ID: 31252) with a total on-source exposure time of 26.37 ks, divided into four observations of approximately  6500 s, each performed for eight days (2014 February-March, see also Table~\ref{tab:log}).  We used two  of the three instruments on board of the {\sl Swift} gamma-ray burst explorer  \citep[see][]{2004ApJ...611.1005G}:  the X-ray Telescope  \citep[XRT; e.g.][]{2005SSRv..120..165B} and the Ultraviolet/Optical Telescope  \citep[UVOT; e.g.][]{2005SSRv..120...95R}.  

The bservations with the XRT were made predominantly in photon-counting mode (PC).  A light curve and spectrum were extracted within a 25-pixel-radius circle  centered on the maximum of the emission from \sgr, which includes more than 95\% of the source flux.
The background was taken from an annulus of 30-pixel inner radius and 60-pixel outer radius, also centered on the source position. This was done using the software XSELECT version 2.4. Observations with the UVOT were made using  the UVW1, UVM2, and U  filters, centered at 2600, 2246, and 3465 \AA, respectively. The light curve was extracted from a 10-pixel-radius region using the UVOTMAGHIST tool version 1.12.  

Observations of the object performed by {\sl Swift} in 2008 and 2012 and reported by \citet{2013MNRAS.435.2822B} are also included in the analysis for completeness.

\section{ High and Low states}

\sgr\  exhibits strong variability on different timescales. We observed the object regularly for five months. The photometric variability appears to be irregular and does not show any periodic signal at the frequency corresponding to the  orbital period reported by \citet{2010PASP..122.1285T}. However, on a much longer timescale, there might be some cyclical activity.  In Figure\,\ref{fig:lc} we present the observed light curve of the object in the V band. Individual measurements are plotted as tiny dots, and the nightly average magnitudes are denoted by filled pentagons. The system varies with  amplitude  as much as 0.4 mag through the course of each night. The full amplitude of variability reaches 1.5\,mag,  changing from 13.6 at the maximum to 14.8 mag at the minimum.  The maximum to minimum cycles occur approximately every 29 days.  We tentatively modeled the light curve with a sinusoid. Only six cycles were covered  during our observations.

\begin{figure*}[ht]
\setlength{\unitlength}{1mm}
\resizebox{11cm}{!}{
\begin{picture}(100,63)(0,0)
\put (-10,-2)  {  \includegraphics[width=16cm, bb=10 150 590 475, clip]{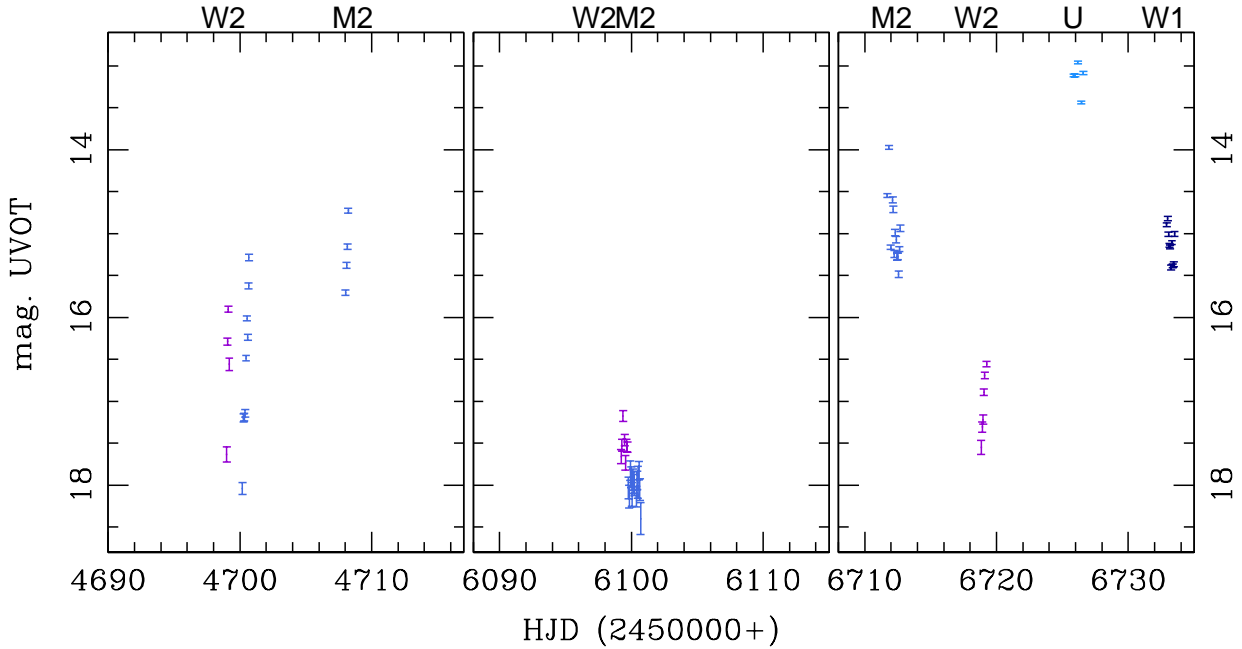}}
\end{picture}}
  \caption{ The {\sl Swift}  UVOT magnitudes of the \sgr.  The object shows strong variability. The UVOT filters used in the observations were set arbitrarily according to the configuration of the instrument at the moment of observation. The filters  are marked on top of the  panels and the data are plotted in different shades of blue to help tell them apart. 
 }
  \label{fig:xuv}
\end{figure*}

The spectral properties of \sgr\ change strongly with the luminosity.  Two distinctive states can be singled out. 
In the high state it has intense emission lines, most notably of the Balmer series accompanied by neutral and ionized helium.  
The Balmer decrement is rather steep (H$_\alpha$\,/H$_\beta$\,/H$_\gamma$\,/H$_\delta$\,/H$_\epsilon$=1.3/1.0/0.85/0.65/0.54). He\,{\sc ii} is ever present in the spectra when it  is in the high state, and  its intensity is comparable to that of H$_\beta$. In addition, a blend of fluorescent lines  of N  {\sc iii}  and C  {\sc iii} around $\lambda 4645$\,\AA\ is prominently present.  Also clearly visible are absorption lines belonging to the secondary star.   In Figure\,\ref{fig:spec} example spectra of the object in the high, the intermediate,  and the low states  are presented. As the object becomes fainter, the lines also weaken. However, the low-state spectrum is highly unusual for a CV  because it contains pure radiation from the secondary star.  Spectra resembling an isolated K star with practically no emission lines have been observed in this system on many occasions in different epochs.  In total we have 10 occurrences of  a low state in the JT spectra and one of our own (as a criteria being considered, the equivalent width (EW) of H$_\alpha$ being less than 4.0\AA).  We witnessed the emergence of emission lines of H$_\beta$  and He\,{\sc ii} from one night to another, with no significant change in the continuum levels.

\citet{2013MNRAS.435.2822B}  report that \sgr\ is also a highly variable source also in the UV and X-rays. 
The UV measurements are presented in  Figure\,\ref{fig:xuv}. The data  were taken in four different filters, unfortunately. That does not help in interpreting  the UV light curve or in determining times when the object was in a low state, i.e. when there was no active accretion going on.  
However, there are recurring observations  obtained in the same UVM2  and UVW2 {\sl Swift} UVOT filters.  A comparison among them  and an analysis of the spectral energy distribution (SED) (presented below in Section\,\ref{sec:sed}) show that  the object probably was in a truly low state only during the 2012 observations corresponding to HJD\,2456099-6100.   During that period of time, two exposures (3400 and 6000\,s long) were obtained over two consecutive days, and the source was persistently faint in the UV and  was barely  visible in the  X-ray. The count rate was $\le0.002\, {\mathrm {cts\,s}^{-1}}$.

Since we know that in the low state the donor star dominates  the flux in the optical domain and there is little evidence of ongoing accretion, we might expect that the X-ray flux  is  formed by the K-type star.  
\citet{2013MNRAS.435.2822B}  estimated that \sgr\ in the low state has a flux of $6.1\times10^{-14}\,{\mathrm {erg\,cm}}^{-2}\,{\mathrm s^{-1}}$\  corresponding to  luminosity  $L_{\mathrm x} =7.3\times10^{30}\,\mathrm {erg\,s^{-1}}$  for a distance of 1 kpc. 
This is consistent with the upper limit  of  the  soft (0.1 - 2.0 keV) X-ray luminosities of RS CVn systems, which are generally found in the range of $10^{29} -10^{32}\,\mathrm {erg\,s^{-1}}$ \citep{1992ApJS...82..311D}. 

The average X-ray spectrum of the object in a broad 0.3--100\,keV range is 100 times brighter, reflecting the high state. The spectrum is consistent with a small  X-ray-emitting region having  plasma temperatures  typical of  a magnetically confined accretion flow, like in polars.  The  $M_{\mathrm{wd}}=0.64\pm0.04$\msun\  mass of the white dwarf was fetched from the  fit to the composite  {\sl XMM-Newton}  EPIC and {\sl Swift} BAT spectrum  \citep{2013MNRAS.435.2822B}.

\begin{figure}[!t]
\centering
 \includegraphics[width=8cm,  clip]{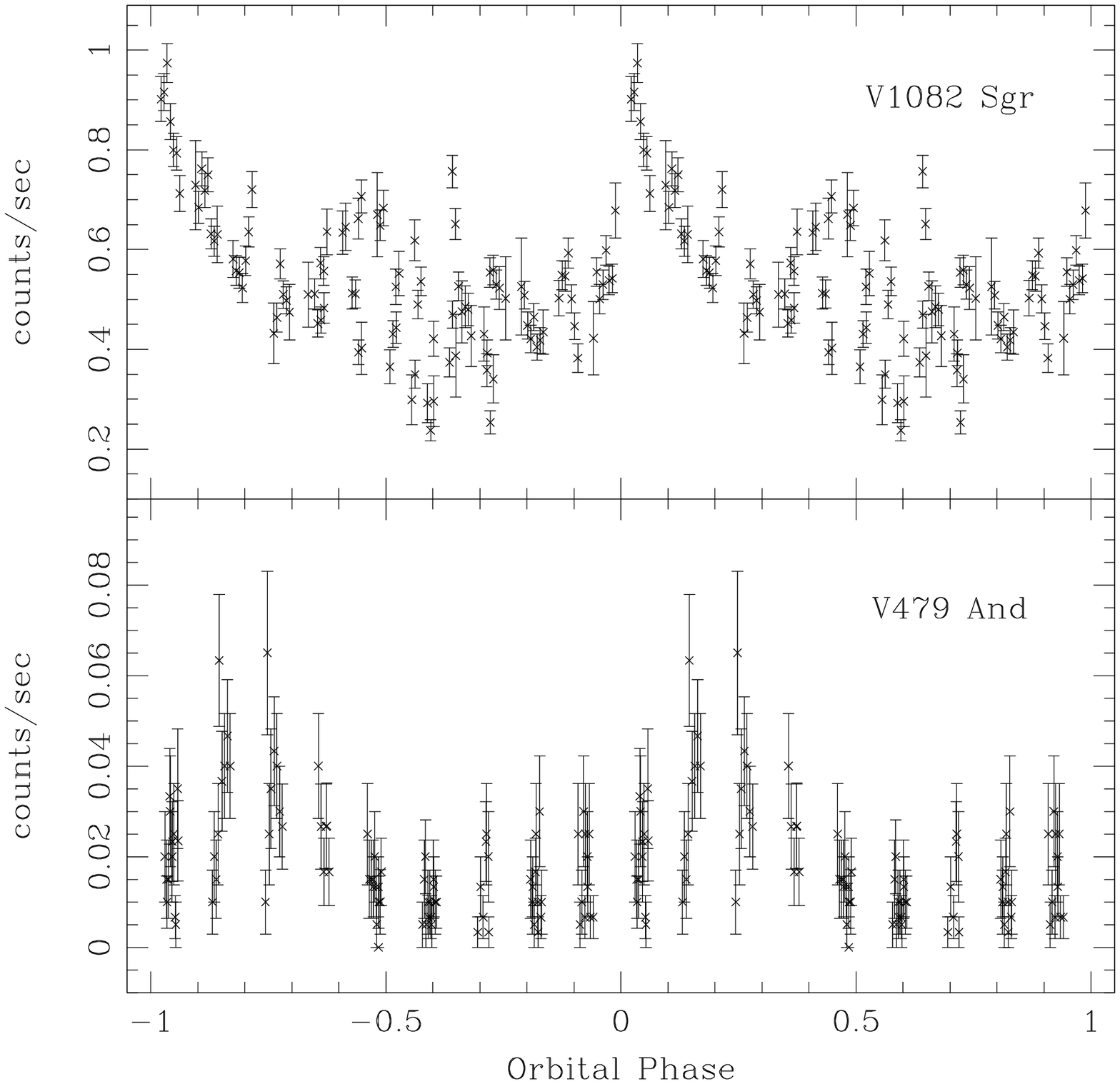}
  \caption{X-ray light curves of \sgr\ and V479\,And folded with their corresponding orbital periods and ephemeris fetched from the absorption-line RVs. The coverage of \sgr\  is not sufficient to prove that the brightness peak is repeating from orbit to orbit, but of  V479\,And it is.  }
  \label{fig:xlc}
\end{figure}

\subsection{Revisiting the X-ray light curve}

\citet{2013MNRAS.435.2822B}  also reported  a prominent brightening , dubbed  as a "flare".  That 5--6 hour flare-like brightening in the X-ray light curve can be  interpreted as  part of an irregular variability, but it also can be the result of a magnetic pole transiting the line of sight. The entire span of time during which {\sl Suzaku} took exposures  of \sgr\ was slightly longer than 30 hr, longer than the orbital period of the object, but not enough to cover two cycles.  If we assume it was not a flare but orbital modulation, then we can fold it  with the $P_{\mathrm {orb}}$ and compare it to V479\,And, a very  similar object contemplated  in this paper. The X-ray light curves of both objects are presented in Figure\,\ref{fig:xlc}, each folded with their corresponding orbital period and ephemeris.  The interesting thing about this plot is that  V479\,And was observed for two continuous orbital cycles, and we know for sure that the brightening there is due to the orbital modulation. In both cases, the X-ray-emitting region is located on the side of the white dwarf facing the donor star and hence comes into the sight of view right after the phase zero. The duration of brightening is similar, and so is the relative flux, although V479\,And is much fainter.  Below we discuss in more detail the similarities of these two objects and their common identity.  

\subsection{Absorption lines}

\begin{figure*}[!t]
\centering
\setlength{\unitlength}{1mm}
\resizebox{11cm}{!}{
\begin{picture}(100,70)(0,0)
%\put (0,0)  { \includegraphics[width=7cm,  clip]{GB_2ndry_spectrum}}
\put (-30,0)  {\includegraphics[width=8cm, clip]{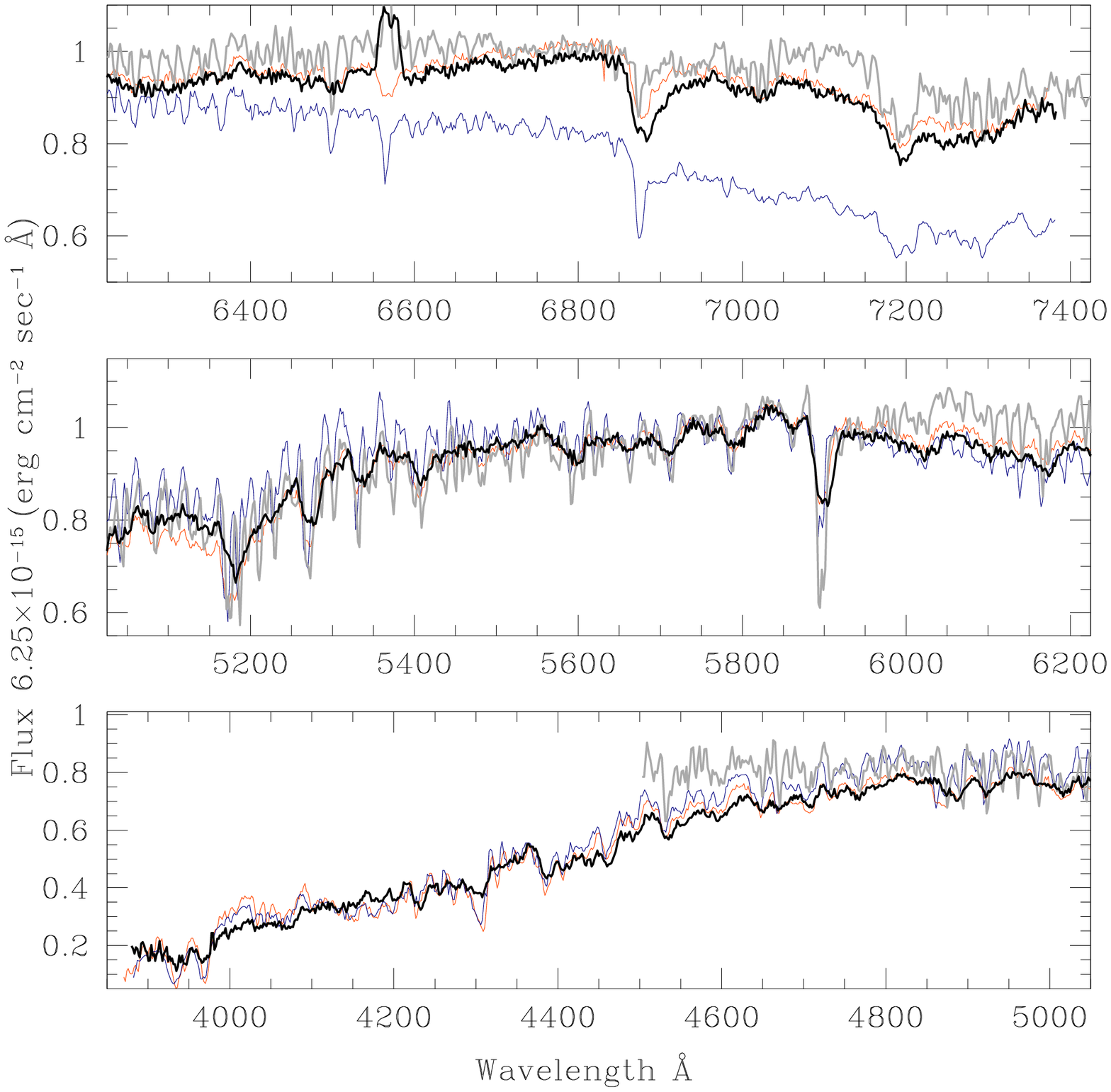}}
 \put (50,0)  {\includegraphics[width=8cm,  clip]{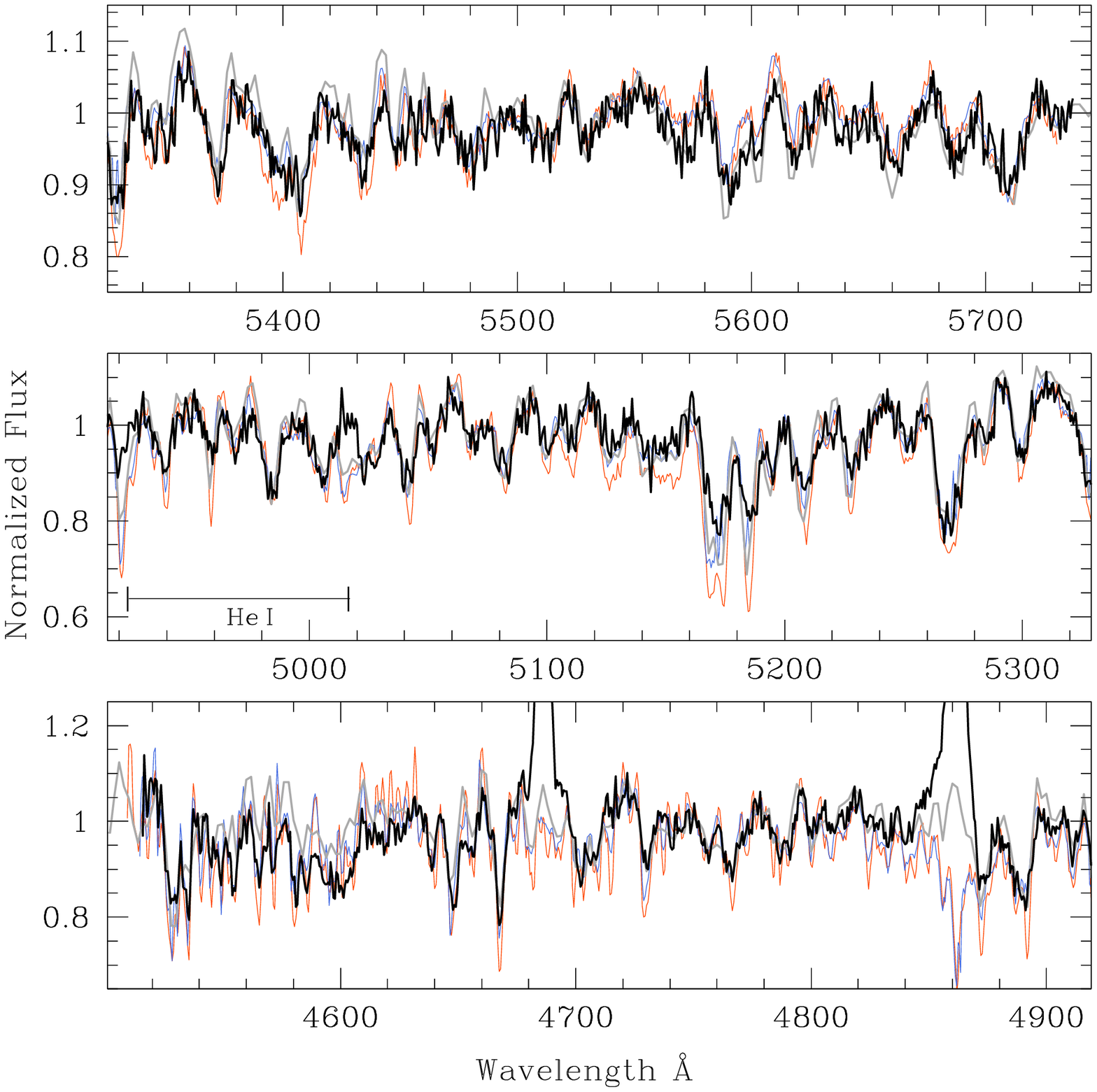}}
\end{picture}}
  \caption{ Details of absorption features of \sgr\  as compared to  comparison stars of known spectral type and luminosity class. The black and gray lines are the spectra of the object from SPM and JT; the latter is scaled in flux to overlap with the  former.  The spectrum of  BSNS104  is represented by  an orange line. The dark-blue line  is the spectrum of HD\,197964, another K2\,IV star classified as such by \citet{2007MNRAS.374..664C}.  They both were observed from SPM at the same night and the same instrumental settings as the object. The depth of lines in the midlle part of the optical range is better matched by HD\,197964, but it deviates at the red end of the spectrum (this might be a flux calibration problem). }
  \label{fig:abslines}
\end{figure*}

\sgr\ experiences low states and, as was mentioned earlier, the spectrum of the object features only the late-type companion of the binary during these episodes.  We only had one chance to observe \sgr\ spectroscopically in a low state (JD=2456893). We used a low-resolution setting and  a wide slit, hence we covered a wide wavelength range with reliable flux calibration.  On the next night, the object already featured emission lines, so it probably was on the path of brightening.  The low-state spectrum is presented in Figures\,\ref{fig:spec} and \ref{fig:abslines}. JT has obtained  14 low-state spectra taken at different epochs. It is not excluded that the flux calibration of their spectra  is not very precise (it is certainly off at the blue end of the spectrum) and the flux is underestimated.  Our spectrum and  the average of JT spectra, scaled to the level of SPM, are displayed in the left panels of Figure\,\ref{fig:abslines}, where large portions of spectra  are plotted in three vertical panels covering the entire observed range.  \citet{2010PASP..122.1285T} suggested a K4 spectral type classification for the donor star of \sgr, but they also mentioned  a large uncertainty of the estimate.
There are many spectra of ZAMS K stars available in catalogs, and one can find a satisfactory continuum flux fit among K2--K4\,V stars, but they fail to match all absorption features. Particularly, the trough corresponding to the MgH band blueward from the Mg\,$b$\  triplet is poorly fit, and the MgH feature around  $ \lambda 4770$\,\AA\  is not compatible with the  main-sequence luminosity class\footnote{A good fit is achieved with K5 star  HD\,283916=SAO\,76803 \citep{1984ApJS...56..257J}, but its spectral classification is disputed by \citet{1997MNRAS.286..500M}, who lists K2 III. Hence we prefer not to rely on this standard.}. We think the K2\,IV spectrum represents a better match to the observed spectra. 

Two selected K2\,IV  stars \citep[BSNS\,104 and HD\,197964;][]{2007MNRAS.374..664C}  scaled to the SPM spectrum are overplotted. They were observed on the same night with the same instrumental settings. BSNS\,104 represents a perfect match, while  HD\,197964 deviates at longer wavelengths. 
However,  HD\,197964 tallies better with the object  than BSNS\,104 in the higher-resolution spectra regarding the depth of absorption features. The higher-resolution spectra obtained on the next night  are presented in the right side panels. Emission lines have already reappeared in \sgr, but the continuum and absorption lines were not affected yet. The spectra in these panels are  normalized. There is a close resemblance of the spectra of the object obtained at SPM, by JT and   HD\,197964. Of course, spectral classification of the donor star should be done with caution, since it may have large spots and, depending on which part of the star surface was observed, the spectral class can vary by two digits.

\begin{figure}[!t]
\centering
 \includegraphics[width=8cm,  clip]{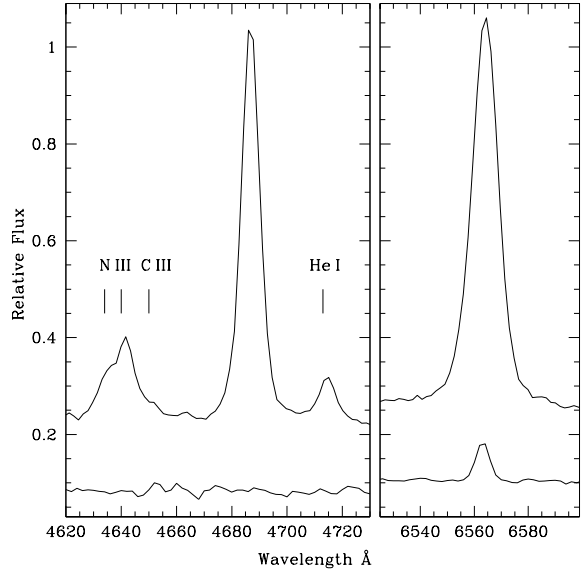}
  \caption{ Example of emission line profiles in a high and low states. In high state the lines are mostly symmetric, single peaked, relatively narrow with wings somewhat wider than a Gaussian profile. They are best described by a single Lorenzian. In the low state H$_\alpha$ is narrow and is not accompanied by other Balmer or helium lines.  }
  \label{fig:eprofiles}
\end{figure}

\begin{figure}[!t]
\centering
 \includegraphics[width=8cm,  clip]{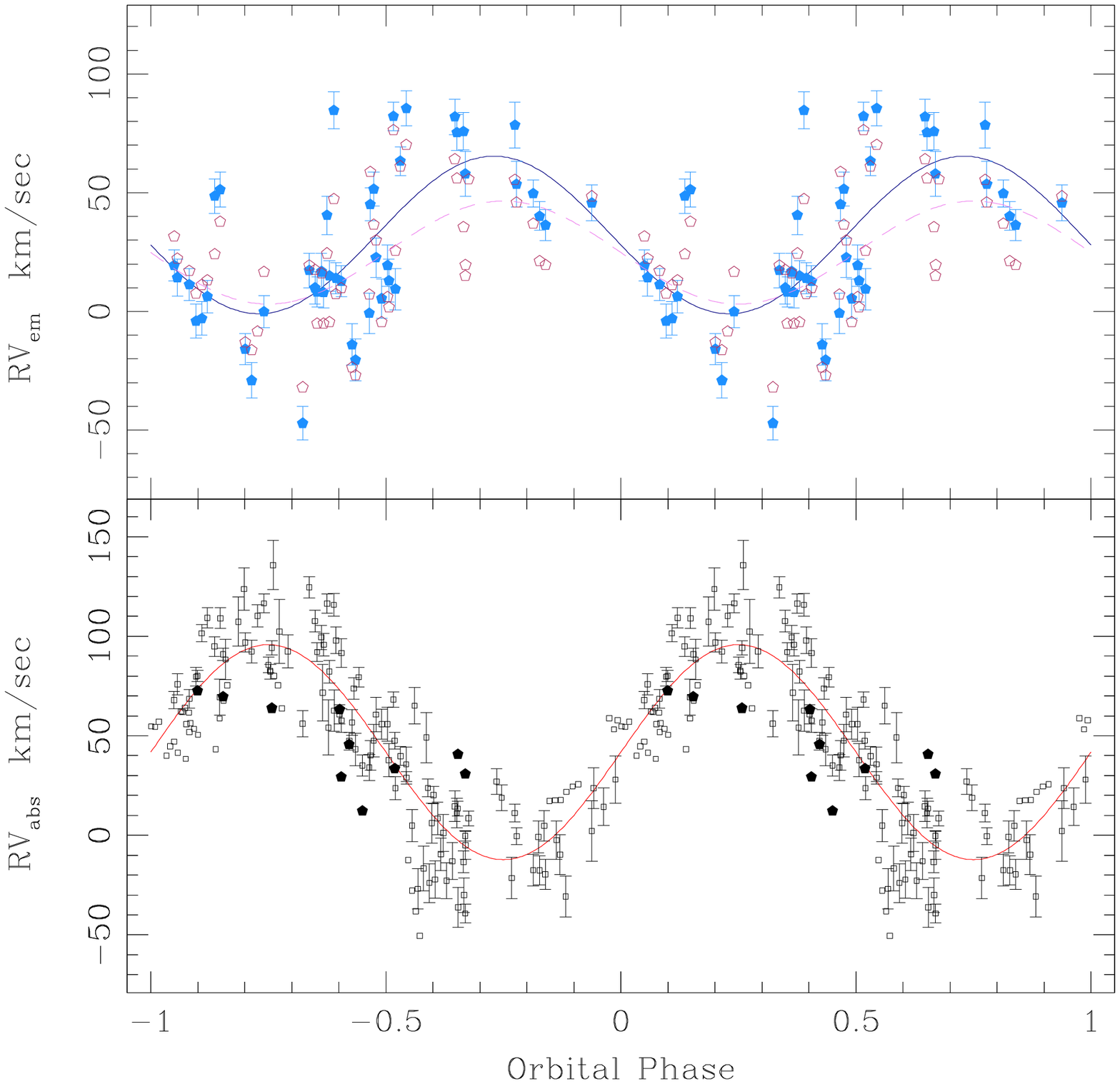}
  \caption{ Radial velocity measurements and sinusoidal fits of different spectral lines. In the bottom panel, the  RVs of a complex of absorption lines are plotted with open squares. Filled pentagons correspond to the RVs of the H$_\alpha$ chromospheric emission  line measured in the low-state spectra.  
In the upper panel, the  measurements of emission lines in higher states are presented.  
The RVs corresponding to H$_\beta$  are plotted with open symbols and a dashed line,  and He\,{\sc ii}   by filled symbols and a continuous line. The data are folded with the orbital period and repeated twice for illustrative purposes.}
  \label{fig:rv}
\end{figure}

The orbital period of \sgr\  was determined by JT using  a complex of absorption features.  We repeated the analysis by adding some reliable RV   measurements obtained by us (adding a longer time base) 
and found no deviation from the period determined by  JT.  We present the absorption lines RV curve in the bottom panel of Figure\,\ref{fig:rv} only for comparison with the  corresponding He\,{\sc ii}  and H$_\beta$\   RV curves.

\subsection{Emission lines}
\label{sec:emlines}

 The emission lines of \sgr\  in a  high state are very intense.  Their intensity is strongly variable, but they remain relatively narrow (FWHM $\approx7-9$\AA; FWZI $\approx40-50$\AA\ of H$_\beta$)  (see Figure\,\ref{fig:spec}).  In  the high state, the EWs of H$_\alpha$\ are in a  $-30$ to $-40$ \AA\ range. 
 In  Figure\,\ref{fig:eprofiles} the profiles of H$_\alpha$  and He\,{\sc ii} are presented. Also included in the plot is a prominent blend of N\,{\sc iii}  and C\,{\sc iii}   around $\lambda 4645$\,\AA. It is formed by a continuum fluorescence, as argued by \citet{1983ASSL..101...97W} and is evidence of a strong UV continuum. The quick resurgence of intense emission lines with little change of continuum and the Balmer decrement asserts that the emission lines are formed in optically thin, sparse gas.   The deficiency of the G-band and N\,{\sc iii}/C\,{\sc iii} profile is probably  evidence of a peculiar chemical composition of the donor star. 

The emission line profiles are single peaked and symmetric. It would be almost impossible to determine the orbital period by them. However, when folded with known orbital periods,  they start to make sense.  A periodic pattern is clearly present, and  it is in  a strict counter-phase from the absorption lines, as demonstrated in Figure\,\ref{fig:rv}.  The measurements of the RVs of H$_\beta$\ and  He\,{\sc ii} are notably dispersed relative to the  best-fit sine curves. The reason is not the accuracy of the measurements, but the intrinsic, chaotic  velocities of the gas superposed on orbital motion.  Nevertheless, both  sine  curves show a strong anticorrelation with the absorption lines formed at the donor star, with the semiamplitude of H$_\beta$  being 30\% less than that of He\,{\sc ii}.  Meanwhile, H$_\alpha$ does not show any periodic variability, just an erratic spread of values. The parameters of the sine fit to H$_\beta$, He\,{\sc ii} and the absorption lines are presented in Table\,\ref{tab:rvs}.

\begin{figure*}[!t]
\centering
\setlength{\unitlength}{1mm}
\resizebox{11cm}{!}{
\begin{picture}(100,50)(0,0)
\put (-20,0)  { \includegraphics[width=16cm,bb = 30 150 650 430,  clip=]{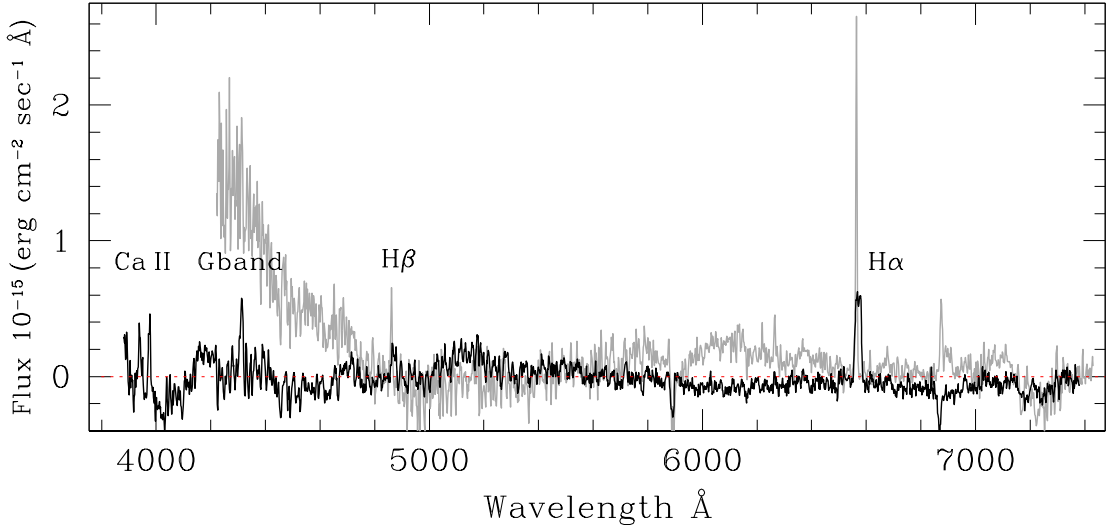}}
\end{picture}}
  \caption{ The residual spectrum  between the  object and  K2,IV star BSNS104 scaled to the continuum at 5500\,\AA is shown. Visible spectral features on the residual spectrum are  H$_\alpha$, a tiny H$_\beta$ and Ca\,{\sc ii} and probably the G band. With a grey line a similar residual spectrum is plotted of the JT spectrum, which has higher spectral resolution but suffers from flux calibration on the blue end of the spectrum. We included it in the plot to demonstrate the narrowness of  H$_\alpha$, leaving no doubts about its chromospheric nature.  }
  \label{fig:2ndry}
\end{figure*}

\begin{table}
 \centering
% \begin{minipage}{140mm}
    \caption{Radial velocity fit parameters.} 
\begin{tabular}{l|cccc} \hline
%&  \multicolumn{4}{c}{Spectroscopy}   &        \\
Line   &   $\gamma$  & RV &   Phase shift  & rms       \\ 
 ID          &    (km s$^{-1}$)         &     (km s$^{-1}$)    &     HJD$_0$  &   (km s$^{-1}$) \\    \hline
Abs.    & $41.8\pm3.5$    & $54\pm4.2$ &  0.0$^\dagger$  & 19.4       \\
H$_\beta$  &   $24.8\pm3.5$    &  $21.7\pm5.5$  &  0.50 &  15.5 \\
He\,{\sc ii}    &   $32.2\pm4.5$    &   $33.2\pm7.5$  &   0.52  &  24.9\\
%He\,{\sc i}   $\lambda 4471$\,\AA   &                                 &                                &  \\
\hline
\end{tabular}
\begin{tabular}{r}
$^\dagger$ {Fixed;  $P_{\mathrm{orb}}$ given by the JT ephemeris }% from \citep{2010PASP..122.1285T}.} 
%\end{minipage}
\end{tabular}
\label{tab:rvs}
\end{table}

In the  low state  we only get to measure  the H$_\alpha$\ line. It has  FWHM $\approx6$\,\AA\ in a  2\AA\,pixel$^{-1}$ resolution spectra.  The  EW in the low state reaches $-2.0$\,\AA, a value very common for chromospherically active K stars \citep{2012MNRAS.421.3189H}.  The RVs  of H$_\alpha$ line with EW$> -4.0$\,\AA\ tend to follow  the motion  corresponding to the secondary.  This is not surprising, assuming that it has a chromospheric origin. The RVs of the H$_\alpha$ line in low-state spectra are marked by filled pentagons in the bottom panel of  Figure\,\ref{fig:rv}. 
In  Figure\,\ref{fig:2ndry} the  residual spectra after  subtraction of a K2\,IV  from the low-state spectrum is presented. The red dashed line corresponds to a zero flux.   Ca\,{\sc ii} H\&K lines become visible in the residual spectrum along  with a  narrow  H$_\alpha$ and are  clearly  of a chromospheric nature. The presence of a faint H$_\beta$\  line  is evidence of enhanced activity.  

\subsection{Spectral energy distribution}
\label{sec:sed}

\begin{figure}[!t]
\centering
 \includegraphics[width=8cm, clip]{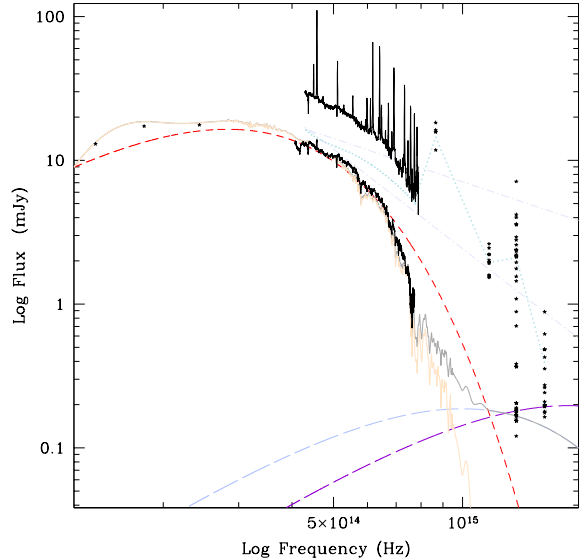}
  \caption{ Spectral energy distribution of \sgr\ in low and high states presented by dark lines and symbols. 
 All available photometric measurements  in IR (2MASS) and UV ({\sl Swift}) regardless of the luminosity state of the object are plotted with the star symbols. The yellow-red  solid  line is a spectrum of K3\,IV  \citep{1998yCat..61100863P}. The dashed lines are calculated blackbodies scaled to pass through observed points. The red short-dashed line corresponds to a 4800\,K, the light-blue
long-dashed line to 17\,000\,K,  and the violet to 30\,000\,K temperature bb.  The gray line is a sum of K3\,IV and 17\,000\,K bb. The light-green short-dashed line is a rough differential flux between high and low states. }
  \label{fig:sed}
\end{figure}

The Roche lobe size of a K2 star  at a 20\,hr orbital period is  $1.6<$\,R/\rsun\,$<1.8$\, for a range of white dwarf masses from 0.5 to 1.2\msun.  Meanwhile, a
 K2\,V star has a radius $0.83$\rsun,  and if the donor star is still on ZAMS, it would hardly fill even  half of its corresponding Roche lobe.  It will take 20-40 Gyr for a K star to evolve to the size of the Roche lobe.  The shortest distance to \sgr\ would be 550\,pc if the donor is a ZAMS star and 1400\,pc if  it is close to 
filling its Roche lobe, assuming  that the average visual magnitude ($m_V=14.7$) at  the low state emanates entirely  from the donor.

In Figure\,\ref{fig:sed} the SED of the object (black lines and symbols) is presented from the IR to UV.  The spectra, as well as photometric measurements, are corrected for the interstellar reddening by assuming $E(B-V)=0.15$ and a standard $R_V=3.1$ \citep{1998ApJ...500..525S} and are recalculated for other wavelengths using the \citet{1999PASP..111...63F} parametrization.  Worth noting is that the hydrogen column density $N_{\mathrm H}=9\times10^{20}\,{\mathrm {cm}}^{-2}$ used in the X-ray spectral analysis corresponds to the same extinction.

 It is not clear in which luminosity state of the object some photometric data were obtained.  The photometric UV data are apparently available in both states. Particularly,   {\sl Swift} UVOT UVM2 band measurements are stretched over a wide energy range; part of  them around  HJD\,2456717 (see Figure\,\ref{fig:xuv}) are certainly taken during the  low state, i.e. a period of time when the contribution from the accretion-fueled processes is negligible.  During such periods,  we observe a pure spectrum of the late star in the optical domain. It would be natural  to assume that the UV flux at such moments is dominated by  the white dwarf.  The temperature of the white dwarf could not be constrained  from the available data. However, we can make a rough estimate from the energy balance.  A white dwarf should have a radius of $R_{\mathrm{wd}}\approx0.01\,{\mathrm{R_{cool}}}$ if the cool star is close to the main sequence, and  two times smaller if it is about  Roche lobe size.  Accordingly, it must have a temperature ranging from  $T_{\mathrm{eff}}\sim17\,000$ to 30\,000\,K  to provide sufficient luminosity and be observable at the low state. Presented in Figure\,\ref{fig:sed} are blackbodies of 17 and 30 kK effective temperature (long-dashed lines), describing a possible contribution from the white dwarf. The red short-dashed line is a black-body of 4800\,K representing the cool star.  A realistic combination of values for the white dwarf radius and the temperature stemming from $L\sim T_{\mathrm{eff}}^4 R_{\mathrm{wd}}^2 (D/ 1{\mathrm{kpc}})^2$ relation, where  $D$ is distance to the object,  confirms correctness of our assumption  that in the low state we observe purely stellar components.

The light-green, short-dashed line roughly corresponds to the flux difference between the high and low states. Dash-dotted lines are power-law\footnote{None of the  blackbody or power curves are  fitted to the observed data (the quantity and quality do not allow one to perform any meaningful fit) and are presented to illustrate the possible configuration of the binary.} $F_\lambda \sim \lambda^{-\alpha}$ with $\alpha=1$ and 2.33. Most CVs radiate in the UV in the form of a power law  with indexes found within that range \citep[e.g.][]{1999MmSAI..70..547D}.  It is possible to find a family of observed points laying on a single power law, except for the U-band data, which might be elevated by the Balmer jump.  Nevertheless,  some measurements are clearly off from being part of a straight line. The additional emission appearing during high states in  the UV does not fit easily into a black body or a power-law model. This is not substantial evidence, but it is a complementary argument to the ones made  in Section\,\ref{sec:emlines}  that an accretion disk or a stream, common in CVs, is not the source of  the excess radiation.

An alternative can be the  cyclotron radiation  since in  the case of low-rate magnetic accretion it provides  the dominant cooling mechanism.  Such radiation  is clearly observed in prepolars containing M dwarfs as a donor star. Of course, in polars and prepolars with shorter periods  ($P_{orb}\la$8hr), 
 the cyclotron radiation is also observed in the optical domain. However, we could not find any credible  signs of cyclotron lines (wide humps in the continuum) in the optical spectra of \sgr\ in the high state. But it  is natural because in the optical range the donor star  of \sgr\  is at least $\sim100$ times brighter than a M4\,V  contained in prepolars.

\section{Interpretation and phenomenological model}

We have demonstrated above that  \sgr\ in the high state has the spectral appearance of a CV, but there are number of irreconcilable details  that invalidate such an interpretation.  One of the basic properties of  CVs is the condition that the donor star fills its Roche lobe (Roche lobe overflow or RLOF) and the system is semidetached.  This leads to the mass loss through the Lagrangian $L_1$ point and formation of a mass-transfer ballistic stream, which upon arrival to the vicinity of the white dwarf either forms an accretion disk or  becomes coupled with the magnetosphere of the magnetic white dwarf and is channeled to the magnetic pole of the latter.  It is believed that in systems with orbital periods longer than three hours the angular momentum  and subsequently the mass loss are driven by magnetic breaking \citep{1981A&A...100L...7V}. Although the mass transfer rate can fluctuate, and  possibly is the  cause of certain types of variability in the rich diversity of CVs, there are no mechanisms  to suddenly halt the mass transfer while the system complies with RLOF and magnetic breaking conditions. It is even more difficult is to imagine a mass transfer/accretion ceasing and restoring  on a  semiregular basis. In \sgr\ we observe such episodes when the accretion-released radiation vanishes repeatedly. 

At the same time, a simple calculation shows that the Roche lobe size in a binary with 15-20 hr period is at least two times larger than the radius of an early-K main-sequence star, regardless of white dwarf mass. 
Rough estimates show  that it will take at least 18 Gyr for a K0 companion to reach the size of the Roche lobe by accounting for magnetic breaking and  gravitational waves for angular momentum loss (O. Toloza 2015, private communication).
If we had only \sgr\ on our hands, maybe we could consider it as a unique case, but the existence of almost  identical V\,479 And makes this assumption improbable. 

However,  if we assume that the donor stars in \sgr\ and V\,479 And are RLOF (with all arguments against) and these two objects are ordinary CVs, we should surmise standard  accretion schemes, proper to CVs.  There are strong arguments against the accretion via disk. A powerful X-ray emission clearly modulated with orbital period in the case of V\,479 And and similar variability in \sgr\ and intense He\,{\sc ii} are not common features of disk systems.  If emission lines are formed in an accretion disk, then both these objects must be of extremely low inclination angle and should harbor unusually massive white dwarfs,  another improbable assumption discrediting the accretion disk scenario. 

It is more natural  to assume that the  white dwarfs in these two objects are magnetic based on the observed X-ray features.  In magnetic CVs or polars, the emission lines in the absence of a disk are formed in accretion streams and are occasionally  on the irradiated side of the donor star. Such emission lines should have very high radial velocities dephased from stellar components or be in phase with the donor-star RVs \citep{1988MNRAS.232..175M}. We do not observe either. Often, emission lines in polars come from two to three different components; then the observed profiles of the lines are very complex, with several S waves forming beautiful patterns in the traced spectra. This is not the case with either object discussed here. 

So it is safe to say that in these two objects the donor stars most probably do not fill their corresponding Roche lobes, even if they are slightly evolved, and that the accretion flows do not follow familiar paths, and hence they are not qualified to be classified as CVs.  What are the alternatives?

We think that  an analogy  with  shorter-orbital-period prepolars  containing M-type companions helps to address all of the  issues  left unanswered by a CV model.  Prepolars are considered  detached binaries containing  a magnetic white dwarf  accreting at a low rate   (\mdot$ \le 10^{-13}$\, \msun\ yr$^{-1}$)  from the stellar wind of  an M dwarf  \citep{2005ApJ...630.1037S}.
Soon after  their discovery  by \citet{1999A&A...343..157R} they were dubbed low-accretion-rate polars, which clearly recognizes the accretion-rate deficiency and, as a consequence, the cooling taking place exclusively via cyclotron radiation \citep{2002ASPC..261..102S}.  The donor stars in low-rate polars are clearly visible and are identified as M stars.   
Thereafter, it was realized that,  at the observed periods, late-M star companions  do not fill their corresponding Roche lobes. 
So  prepolars are rather detached systems, in which the entire stellar  wind of the donor star is captured by the coupled magnetic field and "siphoned" onto the white dwarf \citep[][and references therein]{2005ApJ...630.1037S}.  

However, the analogy is not straightforward  because  \sgr\  and V479\,And   show intense emission lines most of the time, which are permanently absent in prepolars with late-M  donor stars. The objects discussed here have strong X-ray emission, which is  not observed in their short-period counterparts. \sgr\  and V479\,And do not exhibit cyclotron lines in the optical spectra, for which the prepolars became renowned. Most of the differences in appearance are easily explained:
\begin{enumerate}[(a)]
\item Long-period systems containing slightly evolved  K  stars  lose significantly more mass by stellar wind than do M dwarfs. \citet{2002ApJ...574..412W} have argued that the wind mass loss from rapidly rotating, active K  stars can be about $10^3$ times higher than in their older, slow-rotating counterparts.  \citet{2010MNRAS.402.2609I} estimate \mdot$\approx1.2\times10^{-11}$\msun\,yr$^{-1}$\  for a few K1\,IV stars.     
\item Therefore, binaries with early-K donors  may have accretion rates by an order higher than known prepolars.  As it has been demonstrated by \citet{1994MNRAS.268...61L,2005ASPC..330..137W} a sufficiently high magnetic field would siphon the entire stellar wind from the secondary.  Recently, \citet{2012ApJ...758..123W} described it as a "magnetic bottle"  which we find is an appropriate term to present this phenomenon. 
\item A higher accretion rate will require bremsstrahlung to provide sufficient radiative cooling, while making cyclotron emission less prominent \citep{1979ApJ...234L.117L,2014EPJWC..6403001W}.
\item  A collimated X-ray beam from the magnetic pole will create pulses, one per orbital period  (observed in V479\,And), and will ionize the gas  gathered between stars, giving rise to the  emission spectrum. 
\item Meanwhile, the bright donor star would  conceal the cyclotron spectrum in the optical range. 
\item Emission lines would have low RVs with velocities segregated according to the distance to the source of ionization and single-peaked profiles. The final segment of accretion flow will add a broad base to the emission lines, much like the one observed in regular polars \citep{1977ApJ...212L.121C}.
\end{enumerate}

There is  another important difference between short- and long-period prepolars. In the former,  M dwarfs would not evolve far from the ZAMS in a Hubble time, while late-G K stars have a chance to depart. If, after the first common envelope phase, the separation of the binary reaches a  point when the secondary star fills its Roche lobe, its evolution would be altered  \citep{1959ASPL....8...81K,2013A&ARv..21...59I}. In \citet[][see Figure\,12 therein]{2013A&A...553A..28G}, the  authors assumed that the donor star fills its Roche lobe and calculated possible,  but highly improbable, track of the binary in  the 
$P_{\mathrm {orb}}$  vs. spectral type of the donor diagram.  Identification of a second system with very similar characteristics, which in addition is often found in an accretion shutoff state, invalidates that assumption.

If the donor star is smaller than the Roche lobe, it should proceed to evolve as an isolated  star. 
It is difficult to assess the fraction of Roche lobe volume that   donor stars in \sgr\ and  V479\,And  occupy at the moment. There are indications that they might be slightly evolved, but evidence also lacking  that they are in RLOF state.  Since the  dynamical estimate of high white dwarf mass in both these  systems is  not credible, it is fair to assume that they could be close to an average white dwarf mass \citep{2011A&A...536A..42Z}, or less than one solar mass.  Hence, their mass ratio $M_{\mathrm d}/M_{\mathrm {wd}}\ge1.0$, which places them in a zone of instability in the thermal timescale of mass transfer \citep{2015arXiv150704843G},  if the condition of  RLOF was to be true.
  
If the assumption is correct that these two objects are detached binaries with a strongly magnetic white dwarf companion, then we may  conclude that they imitate the CV appearance but they are not CVs.  Hence, they  elude lists of  detached wd$+$K-star systems, introducing a bias in the statistics \citep{2012MNRAS.419..806R}. So far, attempts to find detached binaries with magnetic  white dwarfs and late-type companions were largely fruitless,  with only one firmly identified by \citet{2013MNRAS.436..241P}. 
The shorter period pre-CVs ($P_{orb} \lesssim 8$hr) with M-dwarf components 
have a distinct optical spectra and can be relatively easy to identify. 
The longer-period systems have a more massive and bright G/K companion. They form emission lines in the high mass transfer rate state and become indistinguishable from ordinary CVs, while in the low state they appear as isolated G or K stars.

\section{Conclusions}

We presented spectral, photometric multiwavelength observations of a 20.82 hr orbital period binary \sgr.  It shows distinct high and low states.  Switching from one state to another occurs quite frequently, probably cyclically.  In the high state it appears as a CV, which assumes that the object is a semidetached binary with the donor star filling its Roche lobe and losing matter to a compact companion. But, in the low state, no signs of any accretion are observable, unlike other CVs.   The optical spectrum  is devoid  of emission lines, except for those arising from chromospheric activity, pertaining  to a K star.  The minimum X-ray flux is consistent  with the coronal emission of rapidly rotating  K stars \citep{1992A&A...264L..31G}, and the minimum UV flux is consistent with a contribution from a white dwarf.  No additional continuum flux is detected. It is not clear how the accretion can be frequently halted from a Roche-lobe-overfilling donor star that is under the stress of magnetic breaking. Therefore we assumed that  \sgr\ is a detached binary.  This conclusion is congruent with evolutionary considerations. As it is shown in \citet{2013A&A...553A..28G},  a semidetached binary emerging from the common envelope phase,  containing a K star and having  orbital period similar to V479\,And, will have a peculiar track and will have a limited life as a CV.  On the other hand, it takes several $10^{10}$ years for an isolated K star to reach the size of the Roche lobe in such a binary. Neither is a probable scenario to provide at least two systems with  characteristics otherwise unique for CVs.

Flux ratio,  profiles,  widths, RV amplitudes, orbital phasing, and composition  of emission lines   
argue against presence of an accretion disk or accretion streams like the ones observed in ordinary CVs  \citep{1980ApJ...235..939W,1993AdSAC...9...75R} or  classic polars \citep[see e.g.][] {2004MNRAS.348..316P}.  
In particular, an assumption that the emission lines form in the vicinity of the white dwarf leads to the conclusion that both \sgr\ and  V479\,And,  harbor a massive white dwarf and have an extremely low inclination angle of orbital planes. 
 
On the other hand, the presence and the characteristics of  X-ray and excess UV radiation  and the appearance of an intense He\,{\sc ii} line leaves little prospect but a magnetic accretion on the white dwarf as a source of high-energy radiation. If the white dwarf has a  magnetic field strength corresponding to  known high-field polars  and the K star is similar to chromospherically active stars, the magnetospheres of stellar components can be coupled. That will help to synchronize the spin period of the white dwarf with the orbital and will  help to collect and siphon stellar wind from the  donor star onto the white dwarf.  The essential difference from polars is that the matter is transferred from the donor to the accreting star not through a ballistic trajectory, but instead by flowing  through the bottleneck formed by the closed magnetic lines between stars. Such a trajectory explains the amplitude, the ratio of radial velocities, and the composition  and profiles of the emission lines. The cyclical nature of high and low states and the practically total cessation of accretion in the case \sgr\ might be the result of differential rotation of the K star.   V479\,And does not show such low states. 

A mass transfer induced by the interaction of magnetospheres of detached binaries also helps to explain the lack  of magnetic white dwarfs in detached close binaries.
Magnetic CVs constitute roughly $\ge16\%$ of the total number of CVs, and $\approx10\%$ of isolated white dwarfs are magnetic \citep{2013MNRAS.432..570P}. However, among 2300 detached white$+$red dwarf close binaries, none are found to contain a strongly magnetic white dwarf \citep{2013MNRAS.433.3398R,2014A&A...570A.107R,2014MNRAS.445.1331L}.  
 \citet {2005AJ....129.2376L,2009JPhCS.172a2040L} wondered where the progenitors of magnetic CVs and  magnetic white dwarfs with detached, nongenerate companions are. They cite prepolars with M companions as likely candidates. \citet{2015SSRv..191..111F} agree with this assessment. The two objects discussed in this paper complement a sample of prepolars  with M-dwarf companion. There is a consensus that discovering such systems is a challenging task. In particular, prepolars with M-companions have extremely low optical and X-ray luminosities and no strong emission features or outbursts to reveal them in surveys. In the case of magnetic detached binaries with an earlier components, like V\,1082 Sgr and V\,479 And, the identification is  difficult because they are confused for nova-like CVs  in a high state and  RS\,CVn objects in a low state. In the active accretion mode, they look like low-inclination CVs, and with periods exceeding a few hours such systems  have little  chance to be  studied in detail. Therefore, their statistically small numbers, compared to magnetic CVs and magnetic isolated white dwarfs, might be simply be a result of observational bias.    
The study of these objects lacks a few important components. First of all, we need to find direct evidence of the magnetic nature of the white dwarfs and preferably measure the strength of the magnetic field. We need better X-ray coverage of their long orbital periods in order to understand accretion process. High signal-to-noise ratio  spectroscopy with high resolution of the donor star is very desirable.

\acknowledgments
DGB is grateful to CONACyT for grants allowing his postgraduate studies. GT and SZ acknowledge PAPIIT grants IN107712/IN-100614 and CONACyT grants 166376; 151858 and CAR 208512 for resources provided toward this research. We want to thank Monica Zorotovich and Odette Toloza for helpful discussions regarding the evolution of PCEBs.  We also grateful to Koji Mukai, Domitilla de Martino and Federico Bernardini for inspiring us to study this object and help in interpretation of X-ray data. We thank the anonymous referee for careful reading of the manuscript and helping to make it more congruent.
The results of the paper are in a part based upon observations acquired at the Observatorio Astron\'{o}mico Nacional in the Sierra San Pedro M\'{a}rtir (OAN-SPM), Baja California, M\'{e}xico.
{\sl Facilities:} OANSPM, PROMPT, SWIFT.


%% in parentheses, after the keyword, but they will not be verified.

{\it Facilities:} \facility{OAN SPM}, \facility{PROMPT}, \facility{SWIFT}


\begin{thebibliography}{}

\bibitem[Bernardini et al.(2013)]{2013MNRAS.435.2822B} Bernardini, F., de Martino, D., Mukai, K., et al.\ 2013, \mnras, 435, 2822 

%\bibitem[Bertelli et al.(2008)]{2008A&A...484..815B} Bertelli, G., Girardi, L., Marigo, P., \& Nasi, E.\ 2008, \aap, 484, 815 


%\bibitem[B{\"o}hm-Vitense(2007)]{2007ApJ...657..486B} B{\"o}hm-Vitense, E.\ 2007, \apj, 657, 486 

%\bibitem[Boyle et al.(1992)]{1992A&AS...95...51B} Boyle, R.~P., Dasgupta, A.~K., Smriglio, F., Straizys, V., \& Nandy, K.\ 1992, \aaps, 95, 51 

\bibitem[Burrows et al.(2005)]{2005SSRv..120..165B} Burrows, D.~N., Hill, 
J.~E., Nousek, J.~A., et al.\ 2005, \ssr, 120, 165 

%\bibitem[Beuermann et al.(1998)]{1998A&A...339..518B} Beuermann, K., Baraffe, I., Kolb, U., \& Weichhold, M.\ 1998, \aap, 339, 518 


\bibitem[Cenarro et al.(2007)]{2007MNRAS.374..664C} Cenarro, A.~J., 
Peletier, R.~F., S{\'a}nchez-Bl{\'a}zquez, P., et al.\ 2007, \mnras, 374, 
664

\bibitem[Cieslinski et al.(1998)]{1998A&AS..131..119C} Cieslinski, D., Steiner, J.~E., \& Jablonski, F.~J.\ 1998, \aaps, 131, 119 

\bibitem[Cowley 
\& Crampton(1977)]{1977ApJ...212L.121C} Cowley, A.~P., \& Crampton, D.\ 1977, \apjl, 212, L121 

\bibitem[de Martino(1999)]{1999MmSAI..70..547D} de Martino, D.\ 1999, 
\memsai, 70, 547 


\bibitem[van Dokkum(2001)]{2001PASP..113.1420V} van Dokkum, P.~G.\ 2001, 
\pasp, 113, 1420 

\bibitem[Drake et al.(1992)]{1992ApJS...82..311D} Drake, S.~A., Simon, T., \& Linsky, J.~L.\ 1992, \apjs, 82, 311 


%\bibitem[Dryomova et al.(2005)]{2005A&A...437..375D} Dryomova, G., Perevozkina, E., \& Svechnikov, M.\ 2005, \aap, 437, 375 


%\bibitem[Eggleton(1983)]{1983ApJ...268..368E} Eggleton, P.~P.\ 1983, \apj, 268, 368 

\bibitem[Ferrario et al.(2015)]{2015SSRv..191..111F} Ferrario, L., de 
Martino, D.,  G\"ansicke, B.~T.\ 2015, \ssr, 191, 111 


\bibitem[Fitzpatrick(1999)]{1999PASP..111...63F} Fitzpatrick, E.~L.\ 1999, \pasp, 111, 63

\bibitem[Ge et al.(2015)]{2015arXiv150704843G} Ge, H., Webbink, R.~F., 
Chen, X., \& Han, Z.\ 2015, arXiv:1507.04843 

\bibitem[Gehrels et al.(2004)]{2004ApJ...611.1005G} Gehrels, N., Chincarini, G., Giommi, P., et al.\ 2004, \apj, 611, 1005 


\bibitem[Gonz{\'a}lez-Buitrago et al.(2013)]{2013A&A...553A..28G} Gonz{\'a}lez-Buitrago, D., Tovmassian, G., Zharikov, S., et al.\ 2013, \aap, 553, A28 

%\bibitem[Griffin et al.(2006)]{2006NewA...11..431G} Griffin, R.~F., Church, R.~P., \& Tout, C.~A.\ 2006, \na, 11, 431 

\bibitem[Gudel(1992)]{1992A&A...264L..31G} Gudel, M.\ 1992, \aap, 264, L31

%\bibitem[Echevarr{\'{\i}}a et al.(2007)]{2007A&A...462.1069E} Echevarr{\'{\i}}a, J., Michel, R., Costero, R., \& Zharikov, S.\ 2007, \aap, 462, 1069 

%\bibitem[Hill et al.(2014)]{2014MNRAS.444..192H} Hill, C.~A., Watson, 
%C.~A., Shahbaz, T., Steeghs, D., \& Dhillon, V.~S.\ 2014, \mnras, 444, 192 

\bibitem[Houdebine(2012)]{2012MNRAS.421.3189H} Houdebine, E.~R.\ 2012, 
\mnras, 421, 3189 

%\bibitem[Hussain et al.(2006)]{2006MNRAS.367.1699H} Hussain, G.~A.~J., Allende Prieto, C., Saar, S.~H., \& Still, M.\ 2006, \mnras, 367, 1699 


\bibitem[Ignace et al.(2010)]{2010MNRAS.402.2609I} Ignace, R., Giroux, 
M.~L., \& Luttermoser, D.~G.\ 2010, \mnras, 402, 2609 

\bibitem[Ivanova et al.(2013)]{2013A&ARv..21...59I} Ivanova, N., Justham, S., Chen, X., et al.\ 2013, \aapr, 21, 59 

\bibitem[Jacoby et al.(1984)]{1984ApJS...56..257J} Jacoby, G.~H., Hunter, 
D.~A., \& Christian, C.~A.\ 1984, \apjs, 56, 257

%\bibitem[Kang et al.(2013)]{2013AJ....145..167K} Kang, Y.-W., Yushchenko, 
%A.~V., Hong, K., Guinan, E.~F., \& Gopka, V.~F.\ 2013, \aj, 145, 167 

%\bibitem[Knigge et al.(2011)]{2011ApJS..194...28K} Knigge, C., Baraffe, I., \& Patterson, J.\ 2011, \apjs, 194, 28 

\bibitem[Kopal(1959)]{1959ASPL....8...81K} Kopal, Z.\ 1959, Leaflet of the 
Astronomical Society of the Pacific, 8, 81 

\bibitem[Lamb \& Masters(1979)]{1979ApJ...234L.117L} Lamb, D.~Q., \& Masters, A.~R.\ 1979, \apjl, 234, L117 

%\bibitem[Lehtinen et al.(2012)]{2012A&A...542A..38L} Lehtinen, J., Jetsu, L., Hackman, T., Kajatkari, P., \& Henry, G.~W.\ 2012, \aap, 542, A38 

\bibitem[Li et al.(1995)]{1995MNRAS.276..255L} Li, J., Wickramasinghe, 
D.~T., \& Wu, K.\ 1995, \mnras, 276, 255 

\bibitem[Li et al.(1994)]{1994MNRAS.268...61L} Li, J.~K., Wu, K.~W., 
\& Wickramasinghe, D.~T.\ 1994, \mnras, 268, 61 

\bibitem[Li et al.(2014)]{2014MNRAS.445.1331L} Li, L., Zhang, F., Han, Q., 
Kong, X., \& Gong, X.\ 2014, \mnras, 445, 1331 


\bibitem[Liebert et al.(2009)]{2009JPhCS.172a2040L} Liebert, J.\ 2009, Journal of 
Physics Conference Series, 172, 012040 


\bibitem[Liebert et al.(2005)]{2005AJ....129.2376L} Liebert, J., 
Wickramasinghe, D.~T., Schmidt, G.~D., et al.\ 2005, \aj, 129, 2376 

\bibitem[Malyuto et al.(1997)]{1997MNRAS.286..500M} Malyuto, V., 
Oestreicher, M.~O., \& Schmidt-Kaler, T.\ 1997, \mnras, 286, 500 

%\bibitem[Martinez-Arnaiz et al.(2011)]{2011hsa6.conf..539M} 
%Martinez-Arnaiz, R.~M., Maldonado, J., Montes, D., Eiroa, C., 
%\& Montesinos, B.\ 2011, Highlights of Spanish Astrophysics VI, 539

\bibitem[Mukai(1988)]{1988MNRAS.232..175M} Mukai, K.\ 1988, \mnras, 232, 
175

\bibitem[Parsons et al.(2013)]{2013MNRAS.436..241P} Parsons, S.~G., Marsh, 
T.~R., G{\"a}nsicke, B.~T., et al.\ 2013, \mnras, 436, 241 


%\bibitem[Pickles(1998)]{1998PASP..110..863P} Pickles, A.~J.\ 1998, \pasp, 
%110, 863  

\bibitem[Pickles(1998)]{1998yCat..61100863P} Pickles, A.~J.\ 1998, VizieR 
Online Data Catalog, 611, 863 

\bibitem[Potter et al.(2004)]{2004MNRAS.348..316P} Potter, S.~B., 
Romero-Colmenero, E., Watson, C.~A., Buckley, D.~A.~H., 
\& Phillips, A.\ 2004, \mnras, 348, 316 

\bibitem[Pretorius et al.(2013)]{2013MNRAS.432..570P} Pretorius, M.~L., 
Knigge, C., \& Schwope, A.~D.\ 2013, \mnras, 432, 570 

\bibitem[Rebassa-Mansergas et al.(2012)]{2012MNRAS.419..806R} 
Rebassa-Mansergas, A., Nebot G{\'o}mez-Mor{\'a}n, A., Schreiber, M.~R., et 
al.\ 2012, \mnras, 419, 806 

%\bibitem[Rebassa-Mansergas et al.(2013)]{2013MNRAS.433.3398R} Rebassa-Mansergas, A., Agurto-Gangas, C., et al.\  2013, MNRAS, 433, 3398

\bibitem[Rebassa-Mansergas et al.(2013)]{2013MNRAS.433.3398R} 
Rebassa-Mansergas, A., Agurto-Gangas, C., Schreiber, M.~R., G{\"a}nsicke, 
B.~T., \& Koester, D.\ 2013, \mnras, 433, 3398



\bibitem[Reimers et al.(1999)]{1999A&A...343..157R} Reimers, D., Hagen, H.-J., \& Hopp, U.\ 1999, \aap, 343, 157 


\bibitem[Ren et al.(2014)]{2014A&A...570A.107R} Ren, J.~J., Rebassa-Mansergas, A., Luo, A.~L., et al.\ 2014, \aap, 570, A107 


\bibitem[Ritter(2012)]{2012MmSAI..83..505R} Ritter, H.\ 2012, \memsai, 83, 
505 

\bibitem[Robinson et al.(1993)]{1993AdSAC...9...75R} Robinson, E.~L., 
Marsh, T.~R., 
\& Smak, J.~I.\ 1993, Advanced Series in Astrophysics and Cosmology, 9, 75 


\bibitem[Roming et al.(2005)]{2005SSRv..120...95R} Roming, P.~W.~A., 
Kennedy, T.~E., Mason, K.~O., et al.\ 2005, \ssr, 120, 95 


%\bibitem[Scharlemann(1982)]{1982ApJ...253..298S} Scharlemann, E.~T.\ 1982, 
%\apj, 253, 298 

\bibitem[Schmidt et al.(2005)]{2005ApJ...630.1037S} Schmidt, G.~D., Szkody, 
P., Vanlandingham, K.~M., et al.\ 2005, \apj, 630, 1037 

\bibitem[Schneider 
\& Young(1980)]{1980ApJ...238..946S} Schneider, D.~P., \& Young, P.\ 1980, \apj, 238, 946 

\bibitem[Schwope et al.(2002)]{2002ASPC..261..102S} Schwope, A.~D., 
Brunner, H., Hambaryan, V., 
\& Schwarz, R.\ 2002, The Physics of Cataclysmic Variables and Related Objects, 261, 102 

%\bibitem[Schwope et al.(2006)]{2006A&A...452..955S} Schwope, A.~D., Schreiber, M.~R., \& Szkody, P.\ 2006, \aap, 452, 955 

\bibitem[Schlegel et al.(1998)]{1998ApJ...500..525S} Schlegel, D.~J., 
Finkbeiner, D.~P., \& Davis, M.\ 1998, \apj, 500, 525

%\bibitem[Simon et al.(1985)]{1985ApJ...295..153S} Simon, T., Fekel, F.~C., 
%Jr., \& Gibson, D.~M.\ 1985, \apj, 295, 153 

%\bibitem[Sion et al.(2004)]{2004AJ....128.1795S} Sion, E.~M., Winter, L., 
%Urban, J.~A., et al.\ 2004, \aj, 128, 1795 

%\bibitem[Sion et al.(2012)]{2012ApJ...751...66S} Sion, E.~M., Bond, H.~E., 
%Lindler, D., et al.\ 2012, \apj, 751, 66 

%\bibitem[Southworth et al.(2012)]{2012IAUS..282..123S} Southworth, J., 
%G{\"a}nsicke, B.~T., \& Breedt, E.\ 2012, IAU Symposium, 282, 123 

%\bibitem[Tautvai{\v s}ien{\.e} et al.(2011)]{2011AN....332..925T} 
%Tautvai{\v s}ien{\.e}, G., Barisevi{\v c}ius, G., Berdyugina, S., Ilyin, 
%I., \& Chorniy, Y.\ 2011, Astronomische Nachrichten, 332, 925 



\bibitem[Thorstensen et al.(2010)]{2010PASP..122.1285T} Thorstensen, J.~R., 
Peters, C.~S., \& Skinner, J.~N.\ 2010, \pasp, 122, 1285 

%\bibitem[Tovmassian et 
%al.(2001)]{2001A&A...380..504T} Tovmassian, G.~H., Greiner, J., Zharikov, S.~V., Echevarr{\'{\i}}a, J., \& Kniazev, A.\ 2001, \aap, 380, 504 


%\bibitem[Tovmassian et al.(2002)]{2002AIPC..637...72T} Tovmassian, G., 
%Orio, M., Zharikov, S., et al.\ 2002, Classical Nova Explosions, 637, 72 

%\bibitem[Uchida 
%\& Sakurai(1982)]{1982mpcb.conf...77U} Uchida, Y., \& Sakurai, T.\ 1982, Magnetospheric Phenomena of Celestial Bodies, 77 


\bibitem[Verbunt 
\& Zwaan(1981)]{1981A&A...100L...7V} Verbunt, F., \& Zwaan, C.\ 1981, \aap, 100, L7 


\bibitem[Warner(1995)]{1995CAS....28.....W} Warner, B.\ 1995, Cambridge 
Astrophysics Series, 28

\bibitem[Webbink 
\& Wickramasinghe(2005)]{2005ASPC..330..137W} Webbink, R.~F., \& Wickramasinghe, D.~T.\ 2005, The Astrophysics of Cataclysmic Variables and Related Objects, 330, 137 

\bibitem[Wickramasinghe(2014)]{2014EPJWC..6403001W} Wickramasinghe, D.\ 
2014, European Physical Journal Web of Conferences, 64, 03001 

\bibitem[Williams(1980)]{1980ApJ...235..939W} Williams, R.~E.\ 1980, \apj, 
235, 939 


\bibitem[Williams 
\& Ferguson(1983)]{1983ASSL..101...97W} Williams, R.~E., \& Ferguson, D.~H.\ 1983, IAU Colloq.~72: Cataclysmic Variables and Related Objects, 101, 97 

\bibitem[Wheeler(2012)]{2012ApJ...758..123W} Wheeler, J.~C.\ 2012, \apj, 
758, 123

\bibitem[Wood et al.(2002)]{2002ApJ...574..412W} Wood, B.~E., M{\"u}ller, 
H.-R., Zank, G.~P., \& Linsky, J.~L.\ 2002, \apj, 574, 412 

\bibitem[Zorotovic et 
al.(2011)]{2011A&A...536A..42Z} Zorotovic, M., Schreiber, M.~R., Gansicke, B.~T.\ 2011, \aap, 536, A42 

\end{thebibliography}
\end{document}